\documentclass[aip,twocolumn,superscriptaddress,showpacs,floatfix]{revtex4-1}

\usepackage{amsmath}
\usepackage{amsfonts}
\usepackage{amssymb}
\usepackage{bm}
\usepackage{tabularx}
\usepackage{graphicx}
\usepackage{color}
\usepackage{txfonts}

\usepackage[utf8]{inputenc}
\usepackage[T1]{fontenc}
\usepackage{mathptmx}

\def\CSMSO{Ca$_{1-x}$Sr$_x$Mn$_{0.85}$Sb$_{0.15}$O$_3$}
\def\CMSO{CaMn$_{0.85}$Sb$_{0.15}$O$_3$}

\def\CMMO{CaMn$_{1-x}$Mo$_x$O$_3$}
\def\CCMO{Ca$_{1-x}$Ce$_x$MnO$_3$}
\def\CLMO{Ca$_{1-x}$La$_x$MnO$_3$}

\def\PCMO{Pr$_{1-x}$Ca$_x$MnO$_3$}
\def\YCMO{Y$_{1-x}$Ca$_x$MnO$_3$}

\def\Mn3+{Mn$^{3+}$}
\def\ReEp{$\epsilon_{\mathrm{r}}'$}
\def\ImEp{$\epsilon_{\mathrm{r}}''$}

\def\Trep{$T_{\epsilon'-\mathrm{peak}}$}
\def\Tres{$T_{\epsilon'-\mathrm{sd}}$}
\def\Ties{$T_{\epsilon''-\mathrm{sh}}$}
\def\Tmk{$T_{M-\mathrm{kink}}$}
\def\Tmi{$T_{M-\mathrm{inc}}$}
\def\Trho{$T_{\rho}$}

\draft % marks overfull lines with a black rule on the right

\begin{document}

%\preprint{}

\title{Glassy dielectric anomaly and negative magneto-capacitance effect in electron-doped Ca$_{1-x}$Sr$_x$Mn$_{0.85}$Sb$_{0.15}$O$_3$} %Title of paper



\author{Haruka Taniguchi}
\email[]{tanig@iwate-u.ac.jp}
%\thanks{}
\author{Hidenori Takahashi}
\author{Akihiro Terui}
\author{Kensuke Sadamitsu}
\author{Yuka Sato}
\author{Michihiro Ito}
\author{Katsuhiko Nonaka}
\author{Satoru Kobayashi}
\author{Michiaki Matsukawa}
\affiliation{Iwate University, Morioka 020-8551, Japan}

\author{Ramanathan Suryanarayanan}
%\affiliation{ICMMO, Universit\'{e} Paris-Sud, 91405 Orsay, France (Retired)}
\affiliation{Universit\'{e} Paris-Sud, 91405 Orsay, France (Retired)}

\author{Nae Sasaki}
\author{Shunpei Yamaguchi}
\author{Takao Watanabe}
\affiliation{Hirosaki University, Hirosaki, 036-8561, Japan}


\date{\today}

\begin{abstract}
Manganites exhibit various types of electronic phenomena, and these electronic characteristics can be controlled by carrier doping.
Herein, we report the dielectric and magnetic properties of electron-doped manganite \CSMSO~($x$ = 0, 0.1, 0.2, and 0.3).
The temperature dependence of the real part of the dielectric constant exhibits a broad and large peak just below the kink temperature of magnetization
and a sharp decrease at lower temperatures, accompanied by an anomaly of the imaginary part.
Furthermore, isovalent Sr substitution enhances the temperature of the dielectric peak by more than 50~K.
Interestingly, the dielectric peak exhibits a negative magnetic-field effect.
For all measured samples, the low-temperature variation of the dielectric constant can be qualitatively explained 
based on the Maxwell-Wagner (MW) model that describes a system composed of grain boundaries and semiconducting grains. 
However, the observed peak and its negative magneto-capacitance effect at high temperatures cannot be reproduced 
by a combination of the MW model and magnetoresistance effect.
The dielectric peak strongly indicates polaronic relaxation in the present system.
These results suggest that polarons form clusters with a dipole ordering and magneto-electric coupling,
which might be consistently understood by the charge-ordering scenario.
\end{abstract}


\maketitle %\maketitle must follow title, authors, abstract and \pacs

\section{Introduction}
Manganites exhibit various electronic phenomena such as charge-orbital ordering, colossal magnetoresistance, and phase separation
because of the competition or cooperation between electron--electron interactions (for example, between ferromagnetic double exchange interaction and antiferromagnetic (AFM) super-exchange interaction)~\cite{Tokura1996}.
Carrier doping is a powerful tool to control the electronic state of manganites.
For example, electron-doped systems \CMMO~and \CCMO~change the ground state from a G-type AFM ordering to a charge/orbital-ordered C-type AFM ordering~\cite{Zeng2001, Maignan2002, Caspi2004, Okuda2010}.
In another electron-doped system \CLMO, dielectric properties were reported~\cite{Cohn2004, Cohn2005}.
With respect to hole-doped systems, \PCMO~and \YCMO~exhibit an interesting dielectric anomaly, 
which is sensitive to a magnetic field and is induced around the charge-ordering (CO) temperature~\cite{Jardon1999, Mercone2004, Freitas2005, Serrao2007, Sahu2009}.

With the aim to identify new phenomena in manganites, we investigated an electron-doped system CaMn$_{1-y}$Sb$_{y}$O$_3$ for $y \leq 0.1$~\cite{Murano2010, Murano2011, Fujiwara2013, Fujiwara2015}.
X-ray photoelectron spectroscopy proved that the valency of Sb is 5+ \cite{Fujiwara2013}.
Thus, the substitution of Sb$^{5+}$ ions for Mn$^{4+}$ sites causes one-electron doping with the chemical formula Ca$^{2+}$Mn$^{4+}_{1-2y}$Mn$^{3+}_y$Sb$^{5+}_y$O$^{2-}_3$ 
accompanied by a monotonic increase of unit-cell volume as a function of $y$.
From the magnetization and AC-susceptibility measurements, a canted AFM ordering is expected below the Neel temperature of ~100~K \cite{Murano2010, Murano2011, Fujiwara2013, Fujiwara2015}.
Interestingly, we revealed magnetization reversal after field-cooling for $0.02 \leq y \leq 0.08$~\cite{Murano2010, Murano2011}probably because the local lattice distortion of MnO$_6$ octahedra induced by the Sb substitution changed the orbital state of the $e_g$ electron of Mn$^{3+}$ and reversed the local easy axis of magnetization.
For $y = 0.05$, the physical and chemical pressure effects were also reported~\cite{Fujiwara2013, Fujiwara2015}.

Just recently, we started studying the dielectric properties of \CMSO~\cite{Taniguchi2018} because manganites provide interesting dielectric materials such as magnetoelectrics and relaxors.
Interestingly, we found a broad and large peak in the temperature dependence of the dielectric constant, which exhibited a negative magnetic-field effect.
Considering the broadness and large value of the peak, \CMSO~might be a dielectric glass.
The magnetic-field-sensitive dielectric constant suggests that \CMSO~is a new magnetoelectric material.
Four cases are known to be the origin of magnetoelectrics: magnetic exchange striction~\cite{Fiebig2002, Lorenz2007}, inverse Dzyaloshinskii-Moriya interaction~\cite{Katsura2005, Sergienko2006}, 
$d$-$p$ hybridization~\cite{Arima2007}, and CO~\cite{Ikeda2000, Efremov2004, Efremov2005, Ikeda2005, Khomskii2006, Nagano2007, Brink2008, Giovannetti2009}.
For \CMSO, where both Mn$^{3+}$ and Mn$^{4+}$ ions exist, the formation of CO is expected suggesting a correlation between the CO and the observed magneto-capacitance effect.
CO-originated magnetoelectrics can be realized under these two conditions:
(1) the condition for macroscopic electric polarization\textemdash where the pattern of CO does not have an inversion symmetry; 
(2) the condition for magneto-electric coupling\textemdash where the CO accompanies a magnetic ordering with a one-to-one correspondence between the pattern of the CO and that of the magnetic ordering. 
In such a case, the direction of the electric polarization is also controlled by a magnetic field through a change in the pattern of magnetic ordering.
%Many magnetoelectrics have been identified in manganites~\cite{Fiebig2002, Kimura2003, Hur2004, Kobayashi2004-ErMn2O5, Kobayashi2004-YMn2O5, Kobayashi2004-TbMn2O5, Kenzelmann2005, Kobayashi2005, Arkenbout2006, Heyer2006, Taniguchi2006, Tokunaga2006, Lorenz2007, Picozzi2007, Choi2008}.
%Magnetoelectrics are also found in partially substituted systems that are inevitably inhomogeneous; for example, in a relaxor Pb(Fe$_{1/2}$Nb$_{1/2}$)O$_3$~\cite{Kleemann2010}.
%Thus, the partially substituted manganite \CMSO~is expected to be a magnetoelectric material.

In this study, to clarify the origin of the magneto-capacitance effect of \CMSO, we compared the dielectric, conducting, and magnetic properties in detail.
Moreover, to determine whether \CMSO~is a dielectric glass,
we investigated the frequency dependence of the dielectric constant of Ca$^{2+}_{1-x}$Sr$^{2+}_x$Mn$_{0.85}$Sb$_{0.15}$O$_3$ ($x$ = 0, 0.1, 0.2, 0.3).
%In the temperature dependence of the dielectric constant, we observed a broad peak in the real part, suggesting dipole ordering.
%Importantly, the value of the peak was remarkably suppressed and its temperature was enhanced by increasing the frequency. This result supports that \CSMSO~is a dielectric glass.
%The peak of \ReEp$(T)$ exhibited a negative magnetic-field effect that cannot be explained by magnetoresistance 
%and occurs at almost the same temperature as a magnetization kink which can be related to a CO.
%One possible scenario which is consistent with these results is that \CSMSO~is a CO-originated magnetoelectric material.
%The temperatures of the dielectric peak and the magnetization kink were enhanced to more than 50~K by the isovalent Sr substitution from $x$ = 0 to 0.3.

\section{Experimental}
Polycrystalline samples of \CSMSO~($x$ = 0, 0.1, 0.2, and 0.3) were prepared by a solid-state reaction method.
The stoichiometric mixtures of CaCO$_3$, SrCO$_3$, Mn$_3$O$_4$, and Sb$_2$O$_3$ powders were calcined in air at 1000$^\circ$C for 48~h.
The products were ground and pressed into disk-like pellets.
The pellets were sintered at 1350$^\circ$C for 48~h.
Single crystalline samples were grown by the floating zone method in Ar atmosphere using the polycrystal of \CMSO.
We performed the quantitative composition analysis using the electron probe micro analyzer (JXA-8500F, JEOL).
The compositions of the obtained polycrystalline samples are the same as those of the mixed samples,
whereas that of single crystalline samples is found to be CaMn$_{0.88}$Sb$_{0.12}$O$_3$.
We also performed X-ray diffraction measurements at approximately 290~K using an Ultima I\hspace{-.1em}V diffractometer (Rigaku) with Cu K$\alpha$ radiation.

We measured the dielectric constant under several frequencies and DC magnetic fields using the parallel mode of an LCR meter (Agilent, E4980A).
Samples were cut into a parallel plate with an area of 3.2 $\times$ 6.0~mm$^2$ and a thickness of 0.7 mm, and Au wires for electric lead were connected by Ag paint (Dupont, 4929N).
To reduce the contact resistance, sample surfaces were polished to be flat using 9~$\mu$m diamond slurry, and Ag paint was heated at 110$^\circ$C for 30~min.
We performed measurements with an AC electric field of 1~V/mm and 10~-~500~kHz under 0~-~5~T (field cooling),
obtained the capacitance $C$ and dielectric loss $\tan\delta$,
and estimated the real and imaginary part of the relative dielectric constant \ReEp~and \ImEp.
To investigate the magnetic properties as well, we measured magnetization using a commercial SQUID magnetometer (Quantum Design, MPMS) under 10~mT after field cooling.
DC resistivity and specific heat were measured using the Physical Property Measurement System (Quantum Design).

\section{Results}
\begin{figure*}[htb]
\begin{center}
\includegraphics[width=7in]{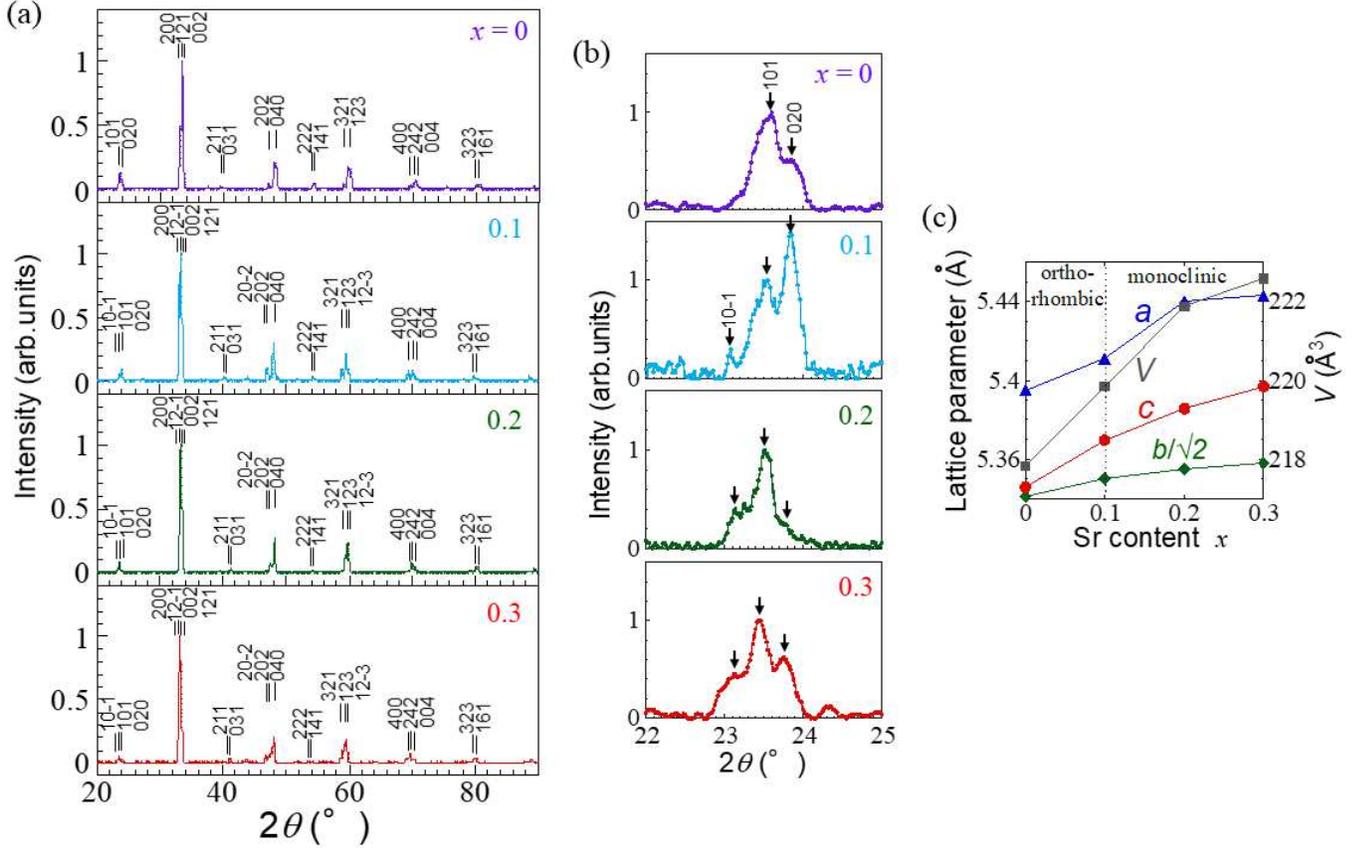}
\end{center}
\caption
{(a) X-ray diffraction spectrum of polycrystalline \CSMSO~($x$ = 0, 0.1, 0.2 and 0.3) normalized for the intensity of the highest peak to be 1.
(b) Enlarged view of (10-1), (101), and (020) peaks. The intensity of the (101) peak is normalized to be 1.
(c) Lattice parameters and unit cell volume of \CSMSO.
}
\label{XRD}
\end{figure*}

Figure~\ref{XRD}~(a) presents the X-ray diffraction spectrum of \CSMSO~($x$ = 0, 0.1, 0.2, and 0.3).
Based on the previous study on orthorhombic CaMn$_{0.9}$Sb$_{0.1}$O$_3$~and monoclinic CaMn$_{0.8}$Sb$_{0.2}$O$_3$~\cite{Poltavets2004},
(101) and (020) peaks are expected at approximately $23.5^\circ$ in both the orthorhombic and monoclinic cases, and (10-1) peak appears at a slightly lower angle in the monoclinic case.
As shown in Fig.~\ref{XRD}~(b), all the samples exhibited a peak at approximately $23.5^\circ$ and $23.8^\circ$, which suggests the (101) and (020) reflection, respectively.
In addition, another peak appeared at approximately $23.1^\circ$ for $x \geq 0.1$, although the peak for $x$ = 0.1 was as small as background fluctuation.
Therefore, we consider that the $x = 0$ sample had a $Pnma$ orthorhombic structure, whereas the $x \geq 0.2$ samples were described by a $P 2_1/m$ monoclinic structure.
The sample of $x = 0.1$ might be located at the boundary between the orthorhombic and monoclinic phases.
These results are similar to the structural variation in other partially substituted manganites \CMMO~and \CCMO~\cite{Maignan2002, Caspi2004, Okuda2010}. 
A stronger intensity at approximately $23.8^\circ$ for $x$ = 0.1, compared with intensity of the other compositions, 
suggests that grinding was imperfect for the powder sample of $x$ = 0.1, and the distribution of the relative angle between the X-ray incidence and the crystallographic axis of the sample was not uniform.
Consistently, the (040) peak at approximately $47.9^\circ$ was also relatively high at $x$ = 0.1, and similarly for the (020) peak at approximately $23.8^\circ$.
The obtained lattice parameters and unit cell volume are plotted in Fig.~\ref{XRD}~(c).
The values of $a$, $b$, and $c$ exhibit a monotonic increase with Sr substitution.
This substitution effect is consistent with the fact that the ion radius of Sr$^{2+}$ is larger than that of Ca$^{2+}$.

\begin{figure*}[htb]
\begin{center}
\includegraphics[width=7in]{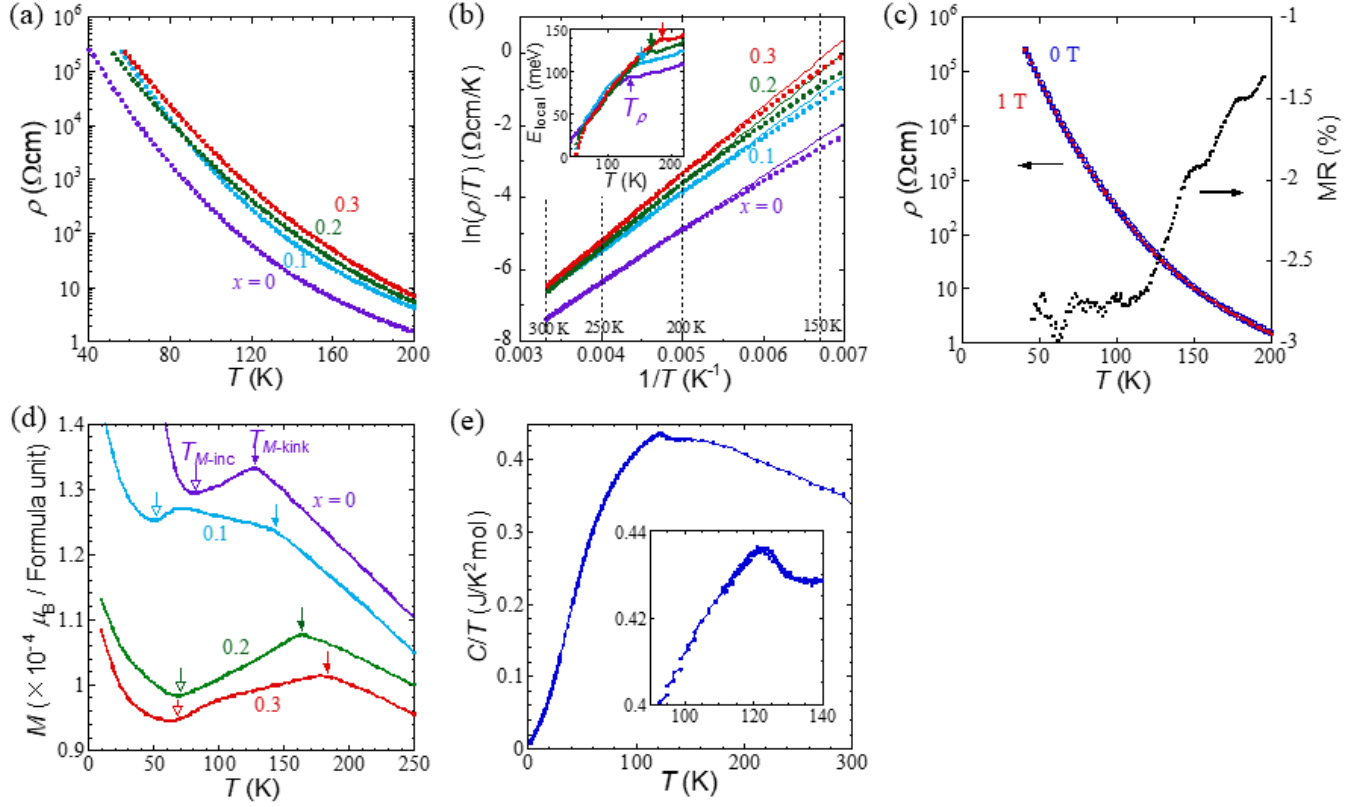}
\end{center}
\caption
{(a) Temperature dependence of the DC resistivity of polycrystalline \CSMSO~($x$ = 0, 0.1, 0.2 and 0.3).
(b) $\ln(\rho/T)$ as a function of $1/T$. The solid lines are the fitting results according to the small polaron hopping model.
The inset presents the local activation energy $E_\mathrm{local} = d(\mathrm{ln}\rho)/d(1/T)$. Arrows designate a small jump-like anomaly that might suggest the formation of short-range CO clusters.
(c) Comparison of the resistivity between 0~T and 1~T for \CMSO.
(d) Temperature dependence of the magnetization measured under 10~mT after field cooling.
Filled and open arrows indicate the positions of the kink and the onset of the remarkable increase, respectively.
(e) Temperature dependence of the specific heat of polycrystalline \CMSO. The inset is an enlarged view near the anomaly.
}
\label{PP}
\end{figure*}

%列と列の間が詰まって見づらいので、空白の列を入れている。
\begin{table*}
\caption{
Characteristic values in the fundamental properties of \CSMSO~($x$ = 0, 0.1, 0.2 and 0.3).
$\rho_{\mathrm{100 K}}$ and $\rho_{\mathrm{200 K}}$ represent the DC resistivities at 100 K and 200 K, respectively. 
The activation energies $E_{\mathrm{a,}\rho}$ are estimated from the high-temperature resistivity data using a small polaron hopping model. 
$T_{\rho}$ denotes the temperature of the small jump-like anomaly in the local activation energy $E_\mathrm{local} (T)$.
$T_{M-\mathrm{kink}}$ denotes the kink temperature of magnetization $M (T)$.
}
\begin{center}
\begin{tabular}{ccccccccccccc} \hline \hline
$x$ && tolerance factor && $\rho_{\mathrm{100 K}}$ && $\rho_{\mathrm{200 K}}$ && $E_{\mathrm{a,}\rho}$ && $T_{\rho}$ && $T_{M-\mathrm{kink}}$ \\
 &&  && ($\Omega$cm) && ($\Omega$cm) && (meV) && (K) && (K) \\ \hline
0.0 && 0.9896 && 295.8 && 1.521 && 127 && 137 && 127 \\
0.1 && 0.9933 && 1640 && 4.262 && 140 && 155 && 143 \\
0.2 && 0.9969 && 2027 && 5.430 && 155 && 169 && 163 \\
0.3 && 1.000 && 3622 && 7.041 && 161 && 189 && 178 \\ \hline \hline
\end{tabular}
\end{center}
\label{T1}
\end{table*}

As shown in Fig.~\ref{PP}~(a), the value of the electrical resistivity $\rho$ of \CSMSO~is smaller than that of typical insulators by several orders of magnitude.
We found that a value of $\rho (T)$ above 190~K can be described by the small polaron hopping model $\rho = \rho_0 T \exp (E_{\mathrm{a,}\rho} / k_\mathrm{B} T)$.
The values of the activation energy $E_{\mathrm{a,}\rho}$ estimated from the fitting shown in Fig.~\ref{PP}~(b) are listed in Table~\ref{T1}.
As shown in the inset of Fig.~\ref{PP}~(b), the temperature dependence of the local activation energy $E_\mathrm{local} = d(\mathrm{ln}\rho)/d(1/T)$ exhibits a small jump-like anomaly.
The fact that both Mn$^{3+}$ and Mn$^{4+}$ ions exist in \CMSO makes CO a reasonable candidate for the transport anomaly.
Because the anomaly is small, CO is expected to be realized only in several fractions in the sample not in the whole sample.
This short-range cluster scenario is supported also by the fact that
substituted Sb ions act as impurities in the Mn lattice as well as dope electrons and suppress the Mn-Mn interaction.
The anomaly temperature \Trho~is listed in Table~\ref{T1}.
As shown in Fig.~\ref{PP}~(c), the magnetoresistance of \CMSO~at 1~T appears to be within the error margin. 

As shown in Fig.~\ref{PP}~(d), the temperature dependence of magnetization $M$ exhibits a kink at a higher temperature 
and a remarkable increase at lower temperatures.
The kink suggests a magnetic order accompanied by a CO because the kink temperature \Tmk~is near \Trho~in \CSMSO.
As shown in Fig.~\ref{PP}~(e), the specific heat of \CMSO~exhibits an anomaly near \Tmk, suggesting phase transition.
By analogy with Ca$_{1-x}$Ce$_x$MnO$_3$~\cite{Caspi2004}, 
in which CO is proved by a neutron diffraction and a similar magnetic kink is observed at the CO temperature~\cite{Zeng2001},
C-type AFM ordering is expected to be formed below \Tmk~in \CSMSO.
The sharp increase in the magnetization below \Tmi~can be understood as the growth of the canted component of AFM ordering.

\begin{figure*}[htb]
\begin{center}
\includegraphics[width=7in]{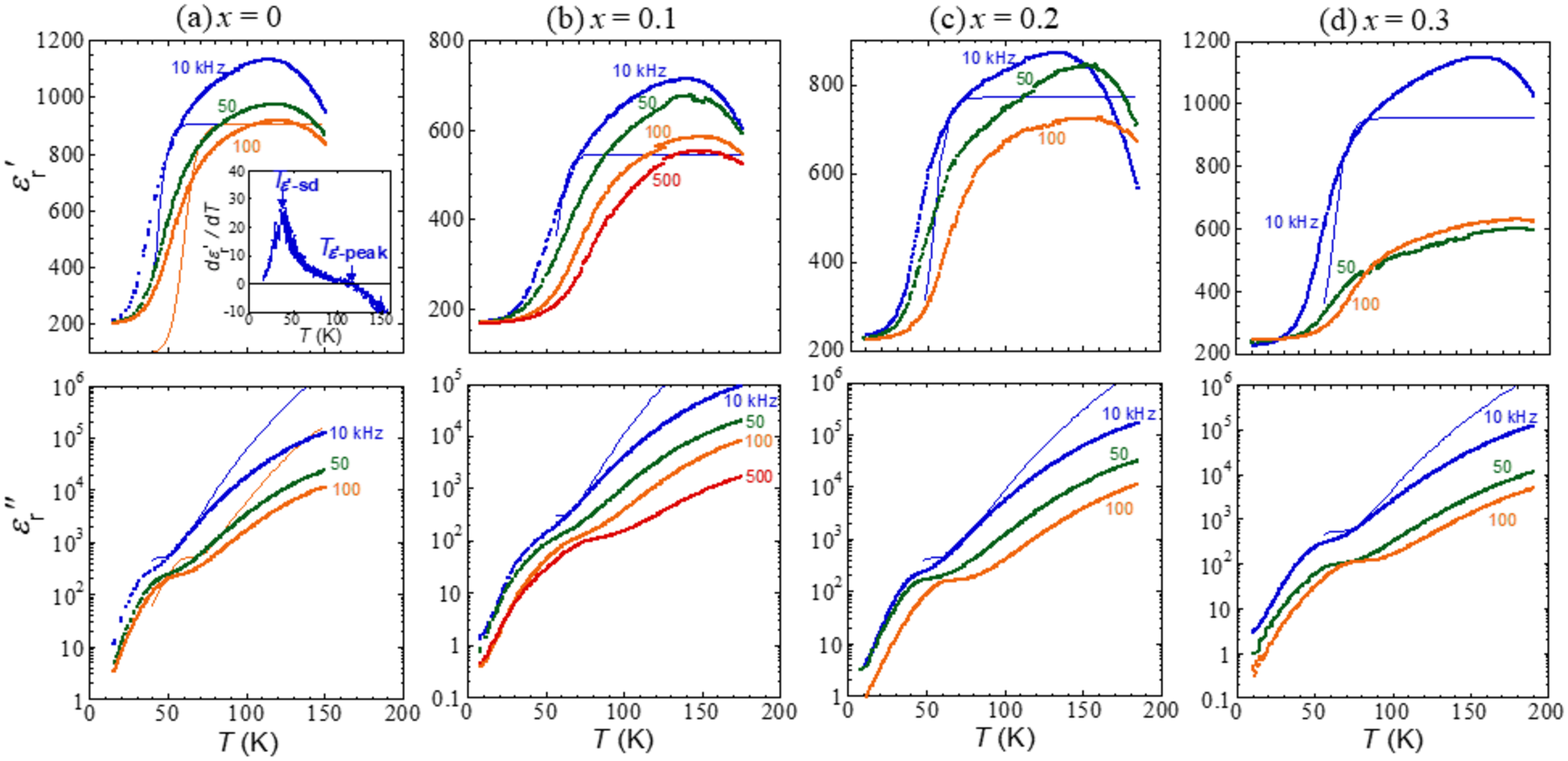}
\end{center}
\caption
{Temperature dependence of the relative dielectric constant of polycrystalline \CSMSO~under several frequencies.
(0~T, AC electric field of 1~V/mm)
(a) $x$ = 0, (b) $x$ = 0.1, (c) $x$ = 0.2, and (d) $x$ = 0.3.
Upper and lower panels represent the real part \ReEp~and the imaginary part \ImEp, respectively.
Each thin solid line designates a Maxwell-Wagner (MW) calculation~\cite{Catalan2006} fitted into each experimental result.
The inset is the temperature dependence of $d\epsilon_{\mathrm{r}}'/dT$.
}
\label{ep-f}
\end{figure*}

In Fig.~\ref{ep-f}, we show the temperature dependence of the relative dielectric constant of \CSMSO~($x$ = 0, 0.1, 0.2 and 0.3) for several frequencies (10, 50, 100 and 500~kHz).
Interestingly, the real part \ReEp~of each sample exhibits a common broad peak at an intermediate temperature 
and then sharply decreases at low temperatures of approximately 50~K. 
Notably, the temperature of the peak in \ReEp$(T)$ is near the temperatures of the anomaly in $E_\mathrm{local}(T)$ and the kink in $M(T)$.
In the imaginary part \ImEp, a shoulder structure appears at approximately 50~K.
% on a temperature dependence which is roughly described by a power law.
For understanding \ImEp($T$), we should note that conductive charges as well as capacitive charges contribute to \ImEp.
Because of the small-polaron-hopping-type temperature dependence and the comparably low value of resistivity,
the dielectric loss of \CMSO~is expected to exhibit an extremely large value at higher temperatures 
and is expected to be remarkably suppressed on cooling.
%the large values of \ImEp~at higher temperatures and its quasi-power-law temperature dependence originate from the current of conductive carriers.
Comparably large \ImEp~was observed in a similar low resistivity system \PCMO~\cite{Shukla2014},
in which \ReEp$(T)$ exhibited a broad peak around the CO temperature and the peak was sensitive to a magnetic field.
Focusing on the frequency dependence, these three anomalies, the peak, the sharp decrease of \ReEp$(T)$, and the shoulder of \ImEp$(T)$, shift to higher temperatures with increasing frequency.
The value of \ReEp~near the peak tends to be suppressed in the high frequency region.

\begin{figure*}[htb]
\begin{center}
\includegraphics[width=7in]{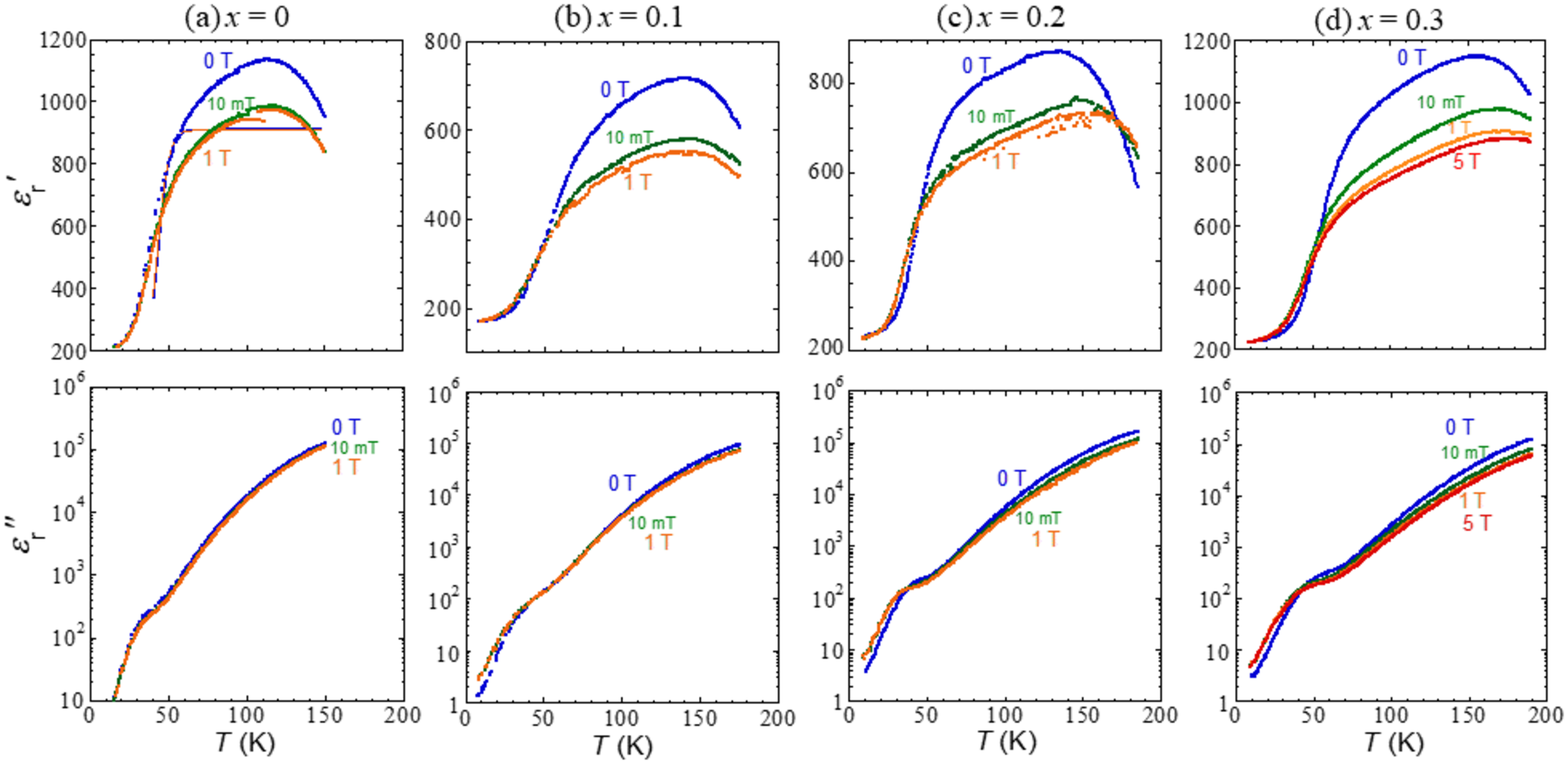}
\end{center}
\caption
{Temperature dependence of the relative dielectric constant of polycrystalline \CSMSO~under several magnetic fields after field cooling.
(AC electric field of 1~V/mm and 10~kHz)
(a) $x$ = 0, (b) $x$ = 0.1, (c) $x$ = 0.2, and (d) $x$ = 0.3.
In the upper panel of (a), two thin solid lines designate the MW calculation~\cite{Catalan2006} fitted into the result for 0~T and 1~T.
Because the resistivity of \CSMSO~rarely exhibits magnetic-field dependence, the two MW curves of 0~T and 1~T almost overlap.
}
\label{ep-H}
\end{figure*}

Figure~\ref{ep-H} presents the effect of magnetic field on the dielectric properties of \CSMSO~($x$ = 0, 0.1, 0.2, and 0.3).
We found a remarkable negative magneto-capacitance effect on the real part: the peak of \ReEp$(T)$ was shifted by the magnetic field.
The peak height was suppressed in all samples, and the peak temperature was enhanced for $x \geq 0.2$.
The magnitude of the magneto-capacitance effect (\ReEp(1~T) - \ReEp(0~T)) / \ReEp(0~T) at the peak temperature of 0~T was -14.4\%, -23.0\%, -17.1\%, and -22.3\% for $x$ = 0, 0.1, 0.2, and 0.3, respectively.
The changes in the imaginary part seem small.

\begin{figure*}[htb]
\begin{center}
\includegraphics[width=5in]{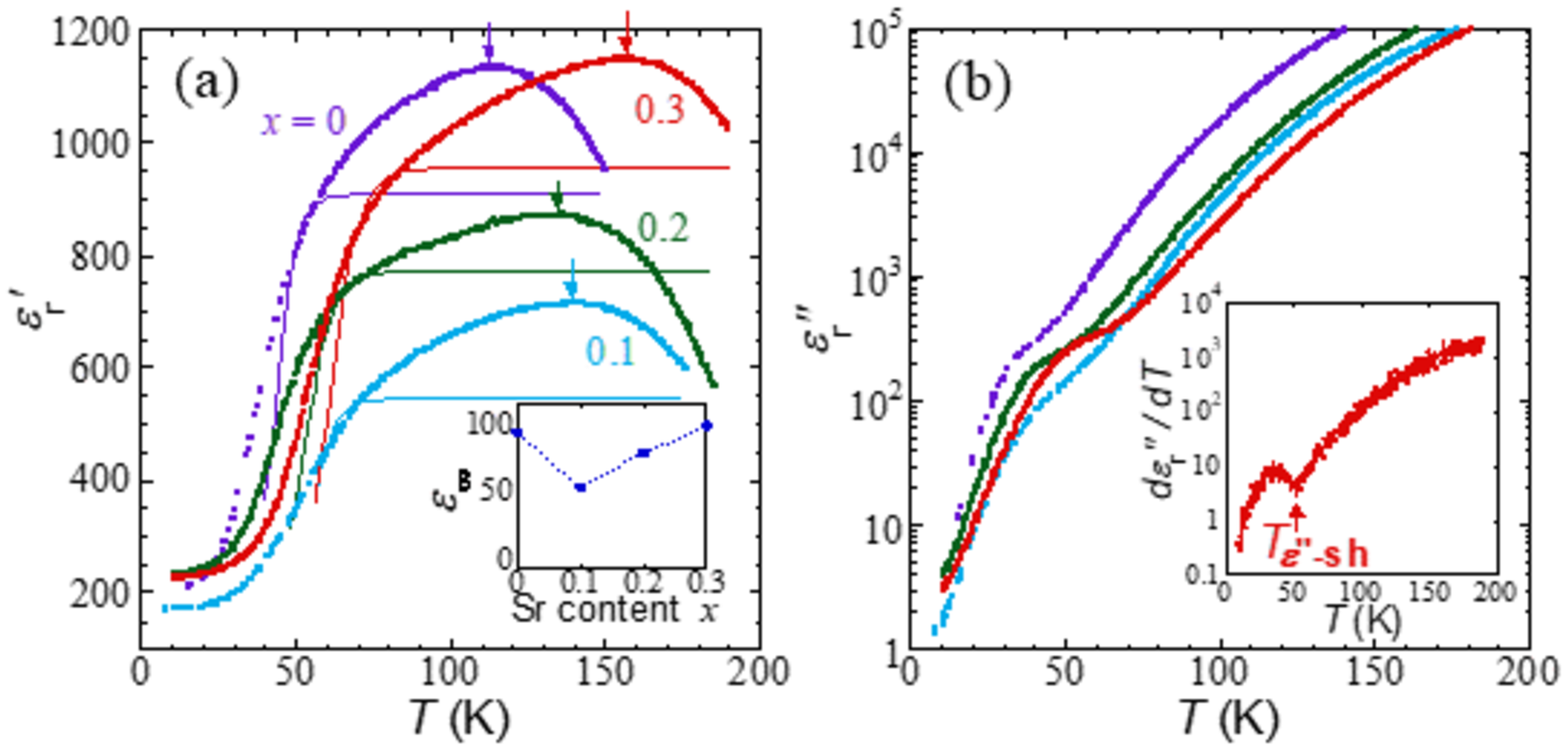}
\end{center}
\caption
{Comparison of the relative dielectric constant among \CSMSO~($x$ = 0, 0.1, 0.2 and 0.3)
under a magnetic field of 0~T and an AC electric field of 1~V/mm and 10~kHz after field cooling.
In panel (a), arrows indicate the peak position and thin solid lines designate the MW calculation~\cite{Catalan2006}.
The inset of (a) is the Sr content dependence of $\epsilon_\infty$ which is a parameter of MW fitting.
The inset of (b) is the temperature dependence of $d\epsilon_{\mathrm{r}}''/dT$ for $x$ = 0.3.
The arrow indicates the minimum temperature, which is defined as the shoulder temperature \Ties.
}
\label{ep-Sr}
\end{figure*}

\begin{figure*}[htb]
\begin{center}
\includegraphics[width=5in]{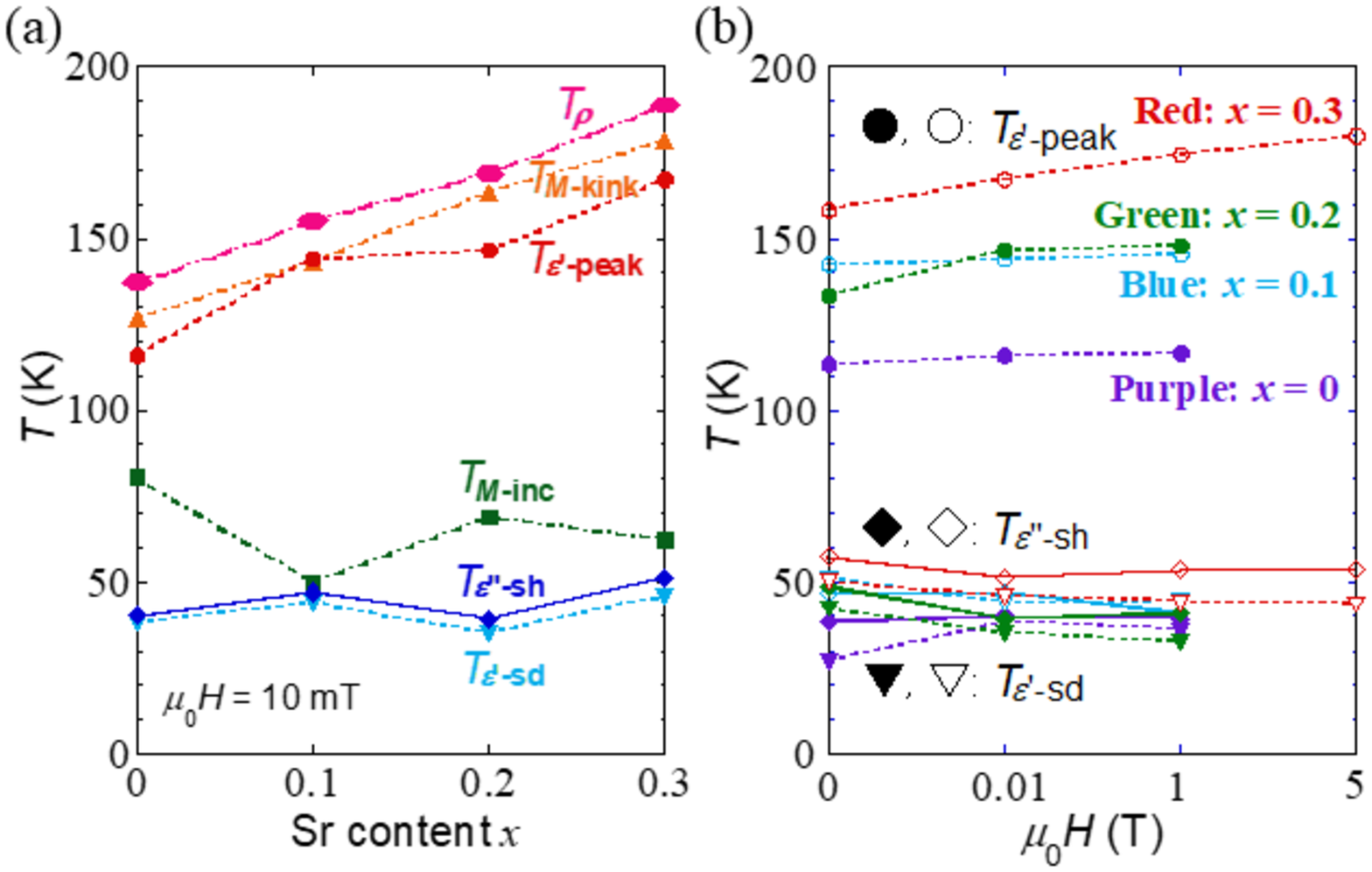}
\end{center}
\caption
{(a) Sr-content and (b) magnetic-field dependence of the anomaly temperatures in the dielectric constant, resistivity, and magnetization of \CSMSO~($x$ = 0, 0.1, 0.2 and 0.3).
Hexagons, triangles, circles, rectangles, diamonds, and reversed triangles indicate \Trho, \Tmk, \Trep, \Tmi, \Ties~and \Tres, respectively.
These anomaly temperatures are defined later in the text.
}
\label{Tc}
\end{figure*}

We estimate the characteristic temperatures of the three dielectric anomalies, one conductive anomaly, and two magnetic anomalies as follows, and plot them in Fig.~\ref{Tc}~(a).
(1) For the peak in the real part of the dielectric constant, the temperature at which $d\epsilon_{\mathrm{r}}'/dT$ becomes zero is defined as the characteristic temperature \Trep.
The positions are shown by arrows in Fig.~\ref{ep-Sr}~(a).
(2) For the sharp decrease in the real part of the dielectric constant,
the temperature at which $d\epsilon_{\mathrm{r}}'/dT (T)$ exhibits a sharp peak is defined as the characteristic temperature \Tres,
an example of which is shown in the inset of Fig.~\ref{ep-f}.
(3) For the shoulder structure in the imaginary part of the dielectric constant, 
the kink temperature at which $d\epsilon_{\mathrm{r}}''/dT (T)$ exhibits the minimum value is defined as the characteristic temperature \Ties,
an example of which is shown in the inset of Fig.~\ref{ep-Sr}~(b).
(4) For the resistivity anomaly,
the temperature \Trho~is determined from the small jump-like anomaly in the local activation energy $E_\mathrm{local}(T) = d(\mathrm{ln}\rho)/d(1/T)$,
as shown by an arrow in the inset of Fig.~\ref{PP}(b).
(5 and 6) For the kink and remarkable increase in magnetization, which are shown by arrows in Fig.~\ref{PP}~(d), 
the characteristic temperatures \Tmk~and \Tmi~are determined from $dM/dT (T)$.
For $x$ = 0.1 and 0.3, another kink is observed at approximately 70~K and 90~K, respectively.
Concerning these additional kinks, we hope to obtain detailed information via specific heat measurements in future work.
To summarize the tendency of the anomaly temperatures, we clarified that \Trep, \Trho, and \Tmk~exhibit a similar value and are enhanced together by Sr substitution ,as shown in Fig.~\ref{Tc}~(a).
Moreover, \Tres~corresponds well to \Ties. They remain at almost the same value under isovalent Sr substitution.

\section{Discussion}
From the broad large-value dielectric peak that exhibits a frequency dependence, \CSMSO~seems to be a dielectric glass.
However, we should carefully examine whether the observed dielectric characters are intrinsic.
A large value of \ReEp~can also be caused by the MW effect
at interfaces like grain boundaries~\cite{Hippel1954, Lunkenheimer2002, Lunkenheimer2010}.
In the MW model, the apparent relative dielectric constant $\epsilon_\mathrm{cal}'$ is described as follows~\cite{Catalan2006}:
\begin{equation}
\epsilon_\mathrm{cal}' (T) = \frac{1}{C_0 (R_1 + R_2)} \frac{\tau_1 + \tau_2 - \tau + \omega^2 \tau_1 \tau_2 \tau}{1 + \omega^2 \tau^2}
\label{ReEp1}
\end{equation}
\begin{equation}
\epsilon_\mathrm{cal}" = \frac{1}{\omega C_0 (R_1 + R_2)} \frac{1 - \omega^2 \tau_1 \tau_2 + \omega^2 \tau (\tau_1 + \tau_2)}{1 + \omega^2 \tau^2}
\label{ImEp1}
\end{equation}
Here, $C_0 = \epsilon_0 S/t$, $\epsilon_0$: electric constant, $S$: sample area, and $t$: sample thickness.
$C_i$ and $R_i$: capacitance and resistance of the corresponding phase, respectively, index $i$ = 1 and 2: grain-boundary layers and semiconducting grains, respectively.
$\tau_1 = C_1 R_1$, $\tau_2 = C_2 R_2$, $\tau = (\tau_1 R_2 + \tau_2 R_1) / (R_1 + R_2)$.
$\omega$: angular frequencies.
Notably, the values of \ReEp~and \ImEp~in this formula are determined by the ratio of the thickness, area, resistivity, and permittivity between the two phases.
According to Catalan's assumption~\cite{Catalan2006}, 
in which $t_{1}/t_{2} = 0.1$, $S_{1} = S_{2}$, $\rho_{1}/\rho_{2} = 100$, and $\epsilon_{1} = \epsilon_{2}$ 
($t_{i}$, $S_{i}$, $\rho_{i}$ and $\epsilon_{i}$: thickness, area, resistivity and intrinsic premittivity of the corresponding phase, respectively,),
Eq.~(\ref{ReEp1}), Eq.~(\ref{ImEp1}), and the relaxation time $\tau$ are transformed as follows:
\begin{equation}
\epsilon_\mathrm{cal}' (T) = \frac{\epsilon_\infty}{10}
\frac{91 + 1000(\epsilon_0 \epsilon_\infty \omega \cdot \rho_2(T))^2}{1 + 100(\epsilon_0 \epsilon_\infty \omega \cdot \rho_2(T))^2}
\label{ReEp2}
\end{equation}
\begin{equation}
\epsilon_\mathrm{cal}" = \frac{1}{10 \omega \epsilon_0 \rho}
\frac{1 + 910(\epsilon_0 \epsilon_\infty \rho \omega)^2}{1 + 100(\epsilon_0 \epsilon_\infty \rho \omega)^2}
\label{ImEp2}
\end{equation}
\begin{equation}
\tau(T) = 10 \epsilon_0 \epsilon_\infty \cdot \rho_2(T)
\label{tau}
\end{equation}
%$\rho$: sample resistivity, 
Here, $\epsilon_\infty \equiv \epsilon_2 / \epsilon_0$.
For fitting Eq.~(\ref{ReEp2}) and Eq.~(\ref{ImEp2}) into the observed \ReEp~and \ImEp,
we adopted the observed DC resistivity values as $\rho _{2}$ of semiconducting grains
because the conducting paths of the majority phase 2 with higher electric conductivity determine the observed resistivity values in the case of the larger volume fraction $t_{1}/t_{2}=0.1$.
As shown in Fig.~\ref{ep-f}, although the sharp decrease of \ReEp~
and the shoulder structure of \ImEp~are reproduced by the MW model, the peak of \ReEp~is not.
Therefore, the peak structure of \ReEp~is expected to be intrinsic.
Moreover, the calculation $\epsilon_\mathrm{cal}'$ does not exhibit a frequency dependence at high temperatures, including \Trep.
This result suggests that the frequency dependence of \ReEp~near \Trep~is also intrinsic.
%Even if the calculated MW effect is subtracted from the observed \ReEp, the peak value seems still large (a few hundred)
%as shown in Table~\ref{T2}.
The value of the fitting parameter $\epsilon_\infty$, which is plotted in the inset of Fig.~\ref{ep-Sr}(a), 
is consistent with that of a similar system La$_{1-x}$Ca$_x$MnO$_3$ (about 20-100)~\cite{Cohn2004}.

%列と列の間が詰まって見づらいので、空白の列を入れている。
\begin{table*}
\caption{
Characteristic values in the dielectric properties of \CSMSO~($x$ = 0, 0.1, 0.2, and 0.3).
Former four values \Trep, $\epsilon'_\mathrm{obs}$($T$=\Trep), $\frac{\epsilon_{\mathrm{obs}}'(1~\mathrm{T})-\epsilon_{\mathrm{obs}}'(0~\mathrm{T})}{\epsilon_{\mathrm{obs}}'(0~\mathrm{T})}$ ($T$=\Trep), and 
\Ties~are estimated from measurement at 10~kHz.
The latter three values $\epsilon'_\mathrm{cal}$(low-$T$), $\epsilon'_\mathrm{cal}$(high-$T$), and $T_0$ are obtained from the calculation based on the MW model~\cite{Catalan2006}.
%In this calculation, we assume that $t_{1}/t_{2}=0.1, \rho _{1}/\rho _{2}=100,$ and $\epsilon _{1}=\epsilon _{2}$, 
%where the parameters $t_{i}, \rho _{i},$ and $\epsilon _{i}$ represent the thickness, the resistivity, and intrinsic dielectric constant of the corresponding phase ($i$=1 and 2: grain-boundary layers and semiconducting grains).
%Here, we adopt the observed d.c. resistivity values as $\rho _{2}$ of semiconducting grains 
%because the conducting paths of the majority phase 2 with higher electric conductivity determine the observed values in the case of the larger volume fraction $t_{1}/t_{2}=0.1$.  
%In the low and high temperature limits for the MW model, we obtain $\epsilon'_\mathrm{cal}$ (Low-$T$)=$\epsilon_\infty$ and  $\epsilon'_\mathrm{cal}$ (high-$T$)=9.1$\epsilon_\infty$. 
%The former and latter cases are responsible for high and low frequency limits, $\omega \tau \gg  1$ and $\omega \tau \ll 1$, respectively.
%($\omega $:the measured frequency,  $\tau $ :the relaxation time of the present MW model)
%Furthermore, we estimate the characteristic temperature $T_0$ of the dielectric loss at which $\omega \tau = 1$. 
%Here, substituting $f$=1 kHz, the vacuum dielectric permittivity, and the fitted low-T limit $\epsilon_\infty$ into $\tau = 10 \epsilon_0 \epsilon_\infty \rho (T)$, 
%we then obtain the resistivity value, giving the corresponding characteristic temperature $T_0$ from the $\rho $($T$) data in Fig.~\ref{PP}~(a). 
}
\begin{center}
\begin{ruledtabular}
\begin{tabular}{ccccccccccccccc}
$x$ && \Trep && $\epsilon'_\mathrm{obs}$($T$=\Trep) && $\frac{\epsilon_{\mathrm{obs}}'(1~\mathrm{T})-\epsilon_{\mathrm{obs}}'(0~\mathrm{T})}{\epsilon_{\mathrm{obs}}'(0~\mathrm{T})}$ ($T$=\Trep) 
&& \Ties && $\epsilon'_\mathrm{cal}$(low-$T$) && $\epsilon'_\mathrm{cal}$(high-$T$) && $T_0$ \\ 
 && (K) &&  && (\%) && (K) && &&  && (K) \\ \hline
0.0 && 113 && 1136 && -14.4 && 38 && 100 && 910 && 43 \\
0.1 && 142 && 717 && -23.0 && 47 && 60 && 546 && 56 \\
0.2 && 133 && 875 && -17.1 && 49 && 85 && 774 && 52 \\
0.3 && 158 && 1151 && -22.3 && 58 && 105 && 956 && 61 \\
\end{tabular}
\end{ruledtabular}
\end{center}
\label{T2}
\end{table*}
% (\ReEp(1~T)-\ReEp(0~T))/\ReEp(0~T)

In the low and high temperature limits for the MW model, we obtained $\epsilon'_\mathrm{cal}$ (Low-$T$) = $\epsilon_\infty$ and $\epsilon'_\mathrm{cal}$ (high-$T$) = 9.1 $\epsilon_\infty$.
At high temperatures, the apparent dielectric constant was enhanced by the reduced effective thickness because of the good conducting grains surrounded by the insulating grain boundaries.
$\epsilon'_\mathrm{cal}$ (Low-$T$) and $\epsilon'_\mathrm{cal}$ (high-$T$) were responsible for the high and low frequency limits, $\omega \tau \gg 1$ and $\omega \tau \ll 1$, respectively.
Table~\ref{T2} 
shows that the characteristic temperatures $T_0$ estimated from the expression of the MW-model-based relaxation agree well with the shoulder temperatures \Ties~for all the measured samples.
Here, $T_0$ is the characteristic temperature of the dielectric loss at which $\omega \tau = 1$. 
For 10~kHz, we obtained the value of $\rho_2 (T_0)$ from Eq.~(\ref{tau})
and estimated the corresponding temperature $T_0$ from the $\rho(T)$ data in Fig.~\ref{PP}~(a).

\begin{figure}[htb]
\begin{center}
\includegraphics[width=2.5in]{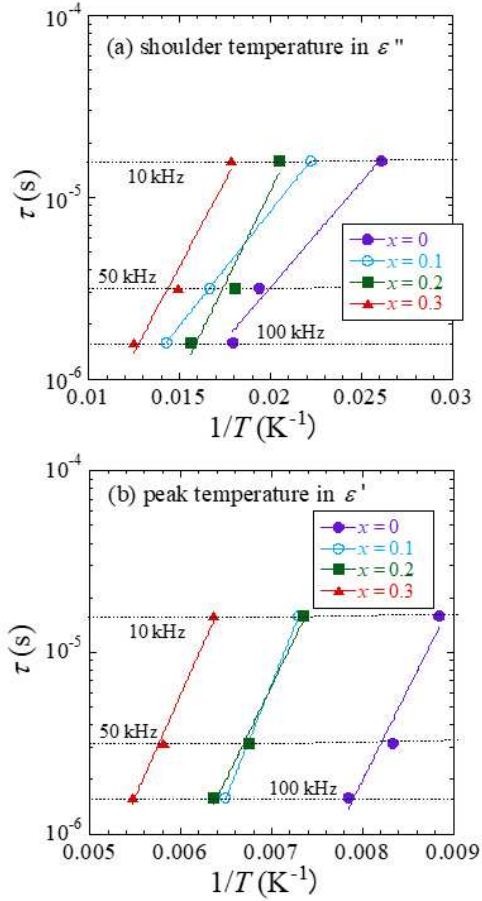}
\end{center}
\caption{
Arrhenius plots from (a) the dielectric-loss shoulder temperature \Ties~and (b) the dielectric peak temperature \Trep~measured at three frequencies ($f$ = 10, 50, and 100~kHz).
}
\label{Aplot}
\end{figure}

In Fig.~\ref{Aplot}, we show the Arrhenius plots from the dielectric-loss shoulder temperature \Ties~and the dielectric peak temperature \Trep~measured at three frequencies ($f$ = 10, 50, and 100~kHz). 
In Table~\ref{T3}, the activation energy $E_{\mathrm{a}}$ and relaxation time $\tau_{0}$ from the Arrhenius plots using $1 / 2 \pi f = \tau = \tau_{0} \exp(E_{\mathrm{a}} / k_\mathrm{B} T)$ are listed as a function of Sr content. 
The parameters, $E_{\mathrm{a}1}$ and $\tau_{01}$, are related to the relaxation process of the heterogeneous material consisting of grain-boundaries and semiconducting grains.
In contrast, $E_{\mathrm{a}2}$ and $\tau_{02}$ estimated from the dielectric peak temperature \Trep~are associated with the relaxation process of localized charges. 
In particular, $E_{\mathrm{a}2}$ and $\tau_{02}$ are comparable to the typical values of polaronic relaxation in CaMnO$_{3}$ and other perovskites~\cite{Cohn2005}. 
The values of $E_{\mathrm{a}2}$ estimated from the dielectric peak are also not very different from the small polaron hopping energies $E_{\mathrm{a,}\rho}$ estimated from the resistivity data.
Therefore, we suggest that the high-temperature dielectric and transport data strongly indicate the polaronic nature of the present system. 

\begin{table}
\caption{
Activation energy $E_\mathrm{a}$ and relaxation time $\tau_{0}$ obtained from the Arrhenius plots using $1 / 2 \pi f = \tau = \tau_{0} \exp(E_{\mathrm{a}} / k_\mathrm{B} T)$,
with $E_{\mathrm{a}1}$ and $\tau_{01}$ being determined from the frequency dependence of the dielectric-loss shoulder temperature \Ties.
$E_{\mathrm{a}2}$ and $\tau_{02}$ were estimated from the dielectric peak temperature \Trep.
}
\begin{ruledtabular}
\begin{tabular}{ccccc}
$x$ & $E_{\mathrm{a}1}$ & $\tau_{01}$ & $E_{\mathrm{a}2}$ & $\tau_{02}$ \\
 & (meV) & (s) & (meV) & (s) \\
\hline
0.0 & 23 & $1.5\times 10^{-8}$ & 198 & $2.1\times 10^{-14}$ \\
0.1 & 25 & $2.5\times 10^{-8}$ & 248 & $1.2\times 10^{-14}$ \\
0.2 & 41 & $8.4\times 10^{-10}$ & 204 & $4.0\times 10^{-13}$ \\
0.3 & 37 & $6.2\times 10^{-9}$ & 264 & $3.0\times 10^{-14}$ \\
\end{tabular}
\end{ruledtabular}
\label{T3}
\end{table} 

\begin{figure}[htb]
\begin{center}
\includegraphics[width=3in]{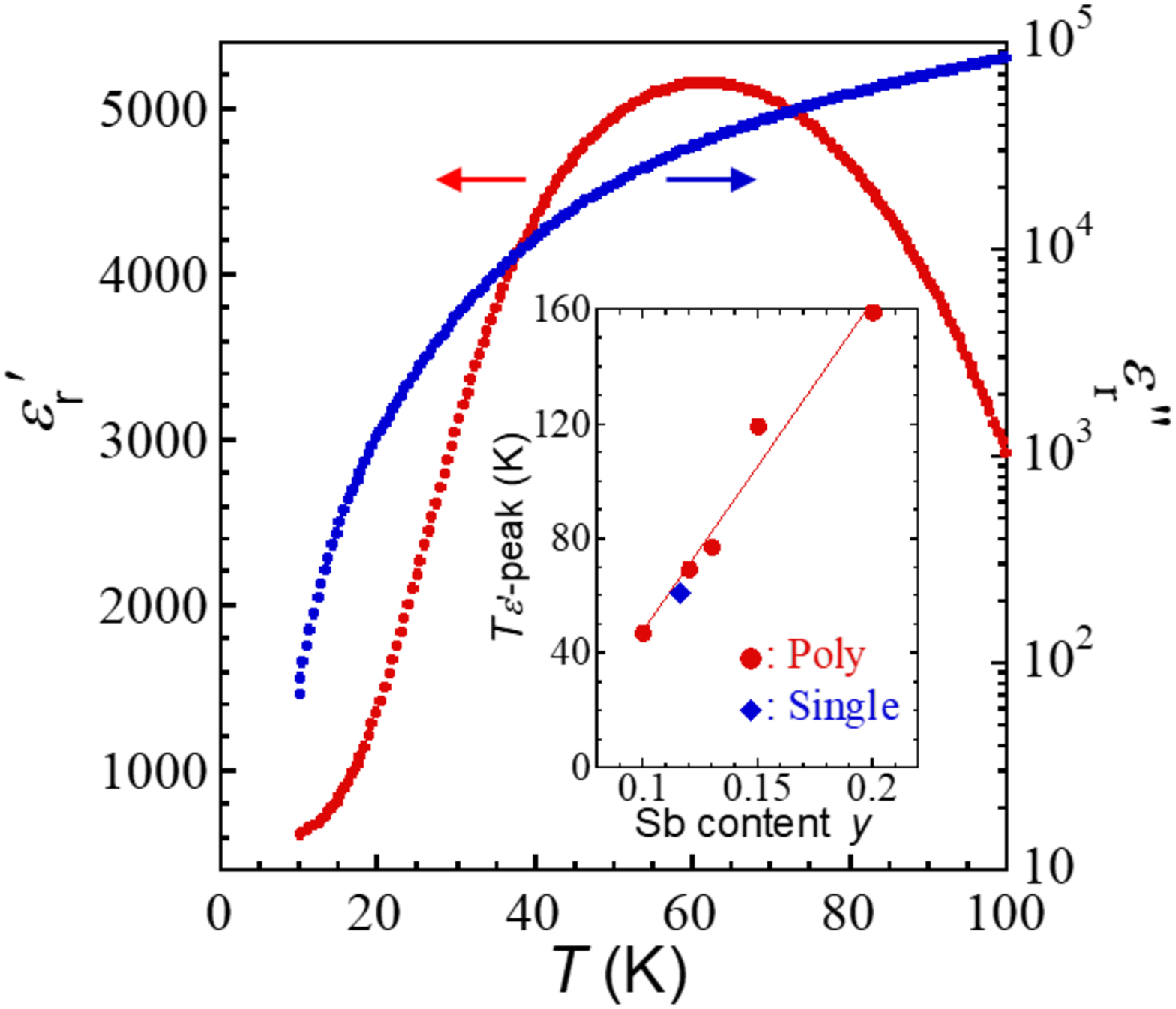}
\end{center}
\caption{
Temperature dependence of the relative dielectric constant of single-crystalline CaMn$_{0.88}$Sb$_{0.12}$O$_3$ under 0~T. (AC electric field of 1~V/mm and 10~kHz)
Inset represents $T_{\epsilon'-\mathrm{peak}}$s of CaMn$_{1-y}$Sb$_{y}$O$_3$.
\Trep~of the single crystal designated by the diamond agreed well with the curve fitted into $T_{\epsilon'-\mathrm{peak}}$s of polycrystals designated by circles~\cite{Taniguchi2019}.
}
\label{Single}
\end{figure}

In addition to the above calculations, we also measured the dielectric constant of 
a single-crystalline CaMn$_{0.88}$Sb$_{0.12}$O$_3$ (thickness: 0.75~mm, cross-sectional area: 13.0~mm$^2$) 
to verify the influence of the MW effect~\cite{Lunkenheimer2010}, as shown in Fig.~\ref{Single}.
The MW effect originating from grain boundaries can be excluded in single-crystalline samples.
In the single crystal, \ReEp$(T)$ exhibits a broad peak at a temperature that is consistent with \Trep~of the polycrystal with the same chemical composition,
as shown in the inset of Fig~\ref{Single}.
This reproducibility of \Trep~indicates that the large-value broad peak of \ReEp$(T)$ in \CSMSO~is not the MW effect but an intrinsic phenomenon.
In contrast, the sharp decrease of \ReEp~and the shoulder of \ImEp~were not observed in the single-crystalline sample, suggesting that these two anomalies at approximately 45~K are caused by the MW effect.
%This result is consistent with the calculation in which the shoulder is suppressed by reducing the interface thickness $t_1$.
%The origin of the sample dependence of \ReEp~value in our case is expected to be the glassy nature of \CSMSO. 
%The sample dependence of the distribution of dielectric domains would result in different \ReEp~value.

Regarding the magnetic-field dependence of \ReEp,
we should note that an apparent effect can be caused by the combination of magnetoresistance and MW effect~\cite{Catalan2006, Kamba2007}.
In addition, from this viewpoint, we can exclude the possibility that the dielectric peak is caused by the MW effect.
When the combination of magnetoresistance and MW effect causes the apparent "magneto-capacitance effect", 
the sign of the "magneto-capacitance effect" changes 
depending on whether the magnetoresistance occurs at the sample core or at the interface~\cite{Catalan2006}. 
First, for the core-dominated magnetoresistance, which is detectable in the high-frequency conductivity, the sign of "magneto-capacitance effect" is positive. 
In contrast, the sign of the observed magneto-capacitance effect is negative in \CSMSO. 
Thus, the origin of the magneto-capacitance effect in \CSMSO~is not core-dominated magnetoresistance. 
Second, for the interface-dominated magnetoresistance, which is detectable in the low-frequency conductivity, 
the magnetoresistance of \CMSO~seems to be within the error margin, as shown in Fig.~\ref{PP}~(c). 
Using the measured resistivity value, 
we calculated the dielectric constant of the MW effect for 0~T and 1~T based on Eq.~(\ref{ReEp2}).
As shown in Fig.~\ref{ep-H}(a), the two calculation curves almost overlapped and could not explain the observed significant magnetic-field effect.
Thus, the origin of the magneto-capacitance effect in \CSMSO~was not interface-dominated magnetoresistance. 
Therefore, we conclude that the magneto-capacitance effect in \CSMSO~which is significant at \Trep~is an intrinsic phenomenon.

From the above discussion, the dielectric characters that are expected to be intrinsic are 
(1) the broad peak structure of \ReEp$(T)$, (2) the frequency dependence of this peak, and (3) the negative magnetic-field effect on this peak.
Because the Arrhenius plots identified the dielectric peak as a phenomenon originating from polarons,
the peak suggests that below the peak temperature \Trep, it becomes difficult for dipole moments consisting of polarons 
to change their direction following the plus-minus switching of an applied AC electric field.
This behavior can be attributed to the spontaneous dipole ordering;
the relaxation time and the activation energy suggest that polarons form the dipole ordering.
In the charge-ordering system \PCMO~whose broad dielectric peak and high conductivity are similar to those of \CSMSO~\cite{Jardon1999, Mercone2004, Serrao2007},
the electric polarization is directly proved by the positive-up-negative-down method~\cite{Shukla2014}.
Thus, by analogy with \PCMO, it is reasonable to expect electric polarization in \CSMSO~below \Trep, at least microscopically.

A broad frequency-dependent peak is commonly observed in dielectric glasses~\cite{Bokov2006}.
Moreover, the ion arrangement of \CSMSO~in which different ions exist on crystallographically equivalent sites 
is favorable for some dielectric glasses.
Therefore, we speculate that \CSMSO~is a dielectric glass consisting of clusters with a short-range dipole ordering:
Because the substituted Sb ions induce disorder (inhomogeneity) into the Mn sites and curb the Mn-Mn interaction,
the dipole ordering caused by an Mn-Mn interaction is of the short-range in \CSMSO.
As a similar example, a spin-glass state is observed in CaMn$_{1-y}$Sb$_{y}$O$_3$ ($0.02 \leq y \leq 0.05$)~\cite{Murano2011}.
%Since some dielectric glasses exhibit ferroelectricity much below the peak temperature of \ReEp$(T)$,
%it is expected to be reasonable for dipole-ordering clusters to grow gradually on cooling.

%Since the response is similar to that of the AC magnetic susceptibility of spin-glass systems, the dipole ordering of \CSMSO~might be glassy.
%Spin-glass behavior in CaMn$_{0.9}$Sb$_{0.1}$O$_3$~\cite{Murano2011} supports the scenario that Ca$_{1-x}$Sr$_{x}$Mn$_{1-y}$Sb$_{y}$O$_3$ has glassy properties.

With respect to coupling between the dielectric and conducting properties,
the similarities in values and the positive correlation in the Sr substitution between \Trep~and \Trho~
suggest that the dielectric peak is related to CO in \CSMSO~($x$ = 0, 0.1, 0.2, and 0.3).
It is expected that the peak of \ReEp$(T)$ is not a secondary anomaly which accompanies changes in conductivity, but it is an intrinsic anomaly originating from capacitive charges.
If a dielectric anomaly is dominated by conductivity, \ImEp~which contains the effect of conductivity as well as the dielectric loss will exhibit a more drastic anomaly than \ReEp. 
However, \ImEp~of \CSMSO~does not exhibit an anomaly at \Trep; 
an example is shown in the inset of Fig.~\ref{ep-Sr}~(b). Even the temperature derivative that can detect a subtle anomaly does not change at \Trep.
Therefore, we consider that the dielectric peak in \CSMSO~is caused by the ordering of localized polarons. CO is a possible example of this ordering.

The magneto-capacitance effect in \CSMSO~can be understood by the hypothesis 
that a spontaneous dipole ordering along a certain direction is stabilized by a magnetic field.
The suppression of the peak height of \ReEp~by a magnetic field suggests that 
the magnetic field fixes the direction of dipole moments in \CSMSO~and disturbs their response to an applied AC electric field.
The enhancement of \Trep~for $x \geq 0.2$ shown in Fig.~\ref{Tc}~(b) indicates that the magnetic field supports a dipole ordering.
%Focusing on the absolute value of the magneto-capacitance effect which is derived from the difference of \ReEp~between zero and non-zero magnetic fields,
%among the manganites whose \ReEp$(T)$ exhibits a glassy broad and large peak,
%the maximum value of about 20\% in \CSMSO~is smaller than that of about 90\% in \PCMO~\cite{Serrao2007} but larger than that of 10\% in \YCMO~\cite{Sahu2009}.
%Compared with other glassy systems CdCr$_2$S$_4$~\cite{Hemberger2005} and Sr$_2$IrO$_4$~\cite{Chikara2009}, 
%it is interesting that the sign of the magneto-capacitance effect is opposite, which suggests that the origin is different.
%Since \Ties~is not almost changed as shown in Fig.~\ref{Tc}~(b), the effect of magnetic field seems to be small on the expansion of polarized dielectric clusters.

The stabilization of the spontaneous dipole ordering along a certain direction by a magnetic field can be explained
by the character of \CSMSO, i.e., the dipole ordering seems to accompany an AFM ordering.
Under 0~T, in a system that has multi crystallographically equivalent axes, 
a magnetic/dipole ordering possesses several possible directions that are degenerated.
A magnetic field introduces anisotropy into the system and fixes the direction of the magnetic ordering.
Because a one-to-one correspondence is expected between the direction of the dipole ordering and that of the AFM ordering,
the direction of the dipole ordering is also fixed by a magnetic field.

Next, we discuss why the substitution of Sr in this compound remarkably increases \Trep, \Trho~and \Tmk~by more than 40~K from $x$ = 0 to 0.3.
Because the Sr$^{2+}$ ion is isovalent to the Ca$^{2+}$ ion, the change of the electronic properties is expected to be caused by a lattice deformation.
As shown in Table~\ref{T1}, the tolerance factor $t = \frac{r_A + r_\mathrm{O}}{\sqrt2 (r_B + r_\mathrm{O})}$ changes from 0.9896 at $x$ = 0 to 1.000 at $x$ = 0.3.
($r_A$, $r_B$, and $r_\mathrm{O}$ are ionic radii of $A$, $B$, and O ion in the $AB$O$_3$ perovskite, respectively.)
The change of $t$ suggests that Mn-O-Mn angles approach 180$^\circ$.
Therefore, we consider that Sr substitution enhances the AFM super-exchange interaction between Mn$^{4+}$ ions 
and assists in the formation of an AFM ordering that is accompanied by a dipole ordering.

As the next issue, it would be important to clarify whether macroscopic electric polarization exists at low temperatures.
Because \CSMSO~exhibits relatively high conductivity, the positive-up-negative-down method would be required~\cite{Traynor1997, Yang2005, Fukunaga2008, Naganuma2008, Shukla2014}.
It would also be interesting to investigate the anisotropy of the dielectric properties using single crystals.

\section{Summary}
We measured the dielectric constant, resistivity, and magnetization of electron-doped manganite \CSMSO~($x$ = 0, 0.1, 0.2, and 0.3).
The temperature dependence of the dielectric constant shows a broad and large peak above 110~K in the real part, 
followed by a sharp decrease at approximately 45~K, at which the imaginary part exhibits a shoulder structure.
The sharp decrease in \ReEp~and the shoulder structure in \ImEp~can be understood by the MW scenario.
The apparent colossal dielectric constant is partially explained by the reduced effective thickness 
due to the good conducting grains surrounded by the insulating grain boundaries at high temperatures.
However, for the following three reasons, the peak in \ReEp~is expected to be intrinsic and suggests a dipole ordering of polarons:
(1) the peak cannot be reproduced by calculations based on the MW model;
(2) the values of the relaxation time and the activation energy that are estimated from the frequency dependence of the peak temperature are typical of polaronic relaxation;
and (3) the peak is observed similarly in single-crystalline samples.
The peak height is suppressed and the peak temperature is enhanced by increasing frequency.
This frequency dependence suggests that \CSMSO~is a dielectric glass.
This hypothesis is reasonable because Sb substitution introduces inhomogeneity in the Mn lattice, and dielectric glasses are often found in such inhomogeneous systems.
Because the temperature of the dielectric peak is similar to that of the resistivity anomaly which suggests a CO,
the dipole ordering in \CSMSO~might be driven by CO.
This dipole ordering accompanies AFM ordering because a magnetization kink was observed near the peak temperature of \ReEp$(T)$.
%One possible CO which causes a spontaneous electric polarization and is accompanied by an AFM ordering
%is the combined state of a site-centered CO and a bond-centered CO.
Notably, we revealed that the dielectric peak exhibits a negative magnetic-field effect, which cannot be explained by magnetoresistance.
We consider that the AFM ordering in \CSMSO~is the glue between the magnetic field and dipole ordering and causes the magneto-capacitance effect.
Substituting Ca$^{2+}$ with isovalent Sr$^{2+}$ remarkably enhances the temperature of the dielectric peak by 50~K
probably because Mn-O-Mn buckling is released and the AFM super-exchange interaction between Mn$^{4+}$ ions is enhanced.

\section*{Acknowledgments}
We acknowledge Y. Ishii, H. Yamamoto, S. Sekikawa, and H. Kimura for fruitful discussions, T. Yajima for the technical support, and H. Fujishiro for the advice on scientific writing.
This work is supported by Iwate University, JSPS KAKENHI Grant Number JP17K14101, and Visiting Researcher's Program of the ISSP.
One of the authors (T. W.) was supported by Hirosaki University Grant for Distinguished Researchers FY2017-2018.
We would like to thank Editage (www.editage.com) for English language editing.

\bibliography{string,Multiferro,Manganite,Others}

\end{document}